\documentclass[article,aps,nofootinbib,showpacs,amsfonts,epsf]{revtex4}

\input{epsf.tex}

\newcommand{\E}{{\cal{E}}}

\renewcommand{\a}{\alpha}
\renewcommand{\k}{\kappa}
\newcommand{\dfrac}[2]{\displaystyle\frac{#1}{#2}}
\newcommand{\be}{\begin{equation}}
\newcommand{\ee}{\end{equation}}
\newcommand{\bea}{\begin{eqnarray}}
\newcommand{\eea}{\end{eqnarray}}
\newcommand{\ba}{\begin{array}}
\newcommand{\ea}{\end{array}}
\def\J#1#2#3#4{{#1} {\bf #2}, #3 (#4)}
\def\PRD{Phys. Rev. D}
\def\PR{Phys. Rev.}
\def\PRL{Phys. Rev. Lett.}
\def\PTP{Prog. Theor. Phys.}
\def\LRR{Living Rev. Relativ.}

\def\AJ{Astrophys. J.}
\def\AL{Astron. Lett.}
\def\MNRAS{Mon. Not. R. Astron. Soc.}
\def\JCAP{J. Cosm. Astropart. Phys.}
\def\JMP{J. Math. Phys.}

\def\CQG{Class. Quantum Grav.}

\def\GRG{Gen. Relativ. Grav.}
\def\PLA{Phys. Lett. A}

\def\ib{{\it ibid.}}

\begin{document}
\draft

\title{The exterior field of slowly and rapidly rotating neutron
stars:\\ Rehabilitating spacetime metrics involving\\
hyperextreme objects}

\author{V. S.~Manko$^\dag$ and E.~Ruiz$\,^\ddag$}
\address{$^\dag$Departamento de F\'\i sica, Centro de Investigaci\'on y de
Estudios Avanzados del IPN, A.P. 14-740, 07000 Cuidad de M\'exico,
Mexico\\ $^\ddag$Instituto Universitario de F\'{i}sica Fundamental
y Matem\'aticas, Universidad de Salamanca, 37008 Salamanca, Spain}

\begin{abstract}
The 4-parameter exact solution presumably describing the exterior
gravitational field of a generic neutron star is presented in a
concise explicit form defined by only three potentials. In the
equatorial plane, the metric functions of the solution are found
to be given by particularly simple expressions that make them very
suitable for the use in concrete applications. Following Pappas
and Apostolatos, we perform a comparison of the multipole
structure of the solution with the multipole moments of the known
physically realistic Berti-Stergioulas numerical models of neutron
stars to argue that the hyperextreme sectors of the solution are
not less (but possibly even more) important for the correct
description of rapidly rotating neutron stars than the subextreme
sector involving exclusively the black-hole constituents. We have
also worked out in explicit form an exact analog of the well-known
Hartle-Thorne approximate metric.
\end{abstract}


\maketitle


\section{Introduction}

In 1995 a general equatorially symmetric two-soliton solution of
the Einstein-Maxwell equations \cite{MMR} (henceforth referred to
as the MMR solution) was constructed as a physically and
astrophysically important application of analytic formulas
determining the extended $N$-soliton metric \cite{RMM} obtained
with the aid of Sibgatullin's integral method \cite{Sib}. Since
then, various particular cases of that solution were analyzed in
the literature in relation with the exterior field of neutron
stars (NSs). Thus, for instance, as an alternative to the
well-known approximate Hartle-Thorne metric \cite{HTh}, in a
series of papers \cite{SSu} Sibgatullin and Sunyaev used for the
description of the NS exterior geometry a 3-parameter vacuum
specialization of the MMR metric \cite{MMRS}, and they
demonstrated that the exact solution with arbitrary parameters of
mass, angular momentum and mass-quadrupole moment was in good
agreement with the numerical models and data from the well-known
paper of Cook, Shapiro and Teukolsky \cite{CST} obtained for
various equations of state (EOS). They also discovered some
universal (independent of the EOS) properties of neutron stars
with regard to the rescaled dimensionless multipole moments. A
limiting case of the MMR solution was considered in the paper
\cite{MMS}, and several authors then studied its 3-parameter
vacuum subcase containing, similar to the solution analyzed by
Sibgatullin and Sunyaev, an arbitrary mass-quadrupole parameter
\cite{SCa,BSt,Han,Bam}. A comparison of the latter subclass with
the numerical models of NSs, performed by Berti and Stergioulas
\cite{BSt} with the help of an advanced numerical code
\cite{SFr,Ste}, revealed in particular that the 3-parameter
analytic solution was better suited for modeling the geometry
around rapidly rotating NSs, giving at the same time a worse
matching with the numerical data for slowly rotating NSs. Even
though afterwards the papers of Pappas and Apostolatos
\cite{PAp1,PAp2} helped to considerably reduce the discrepancies
between the analytical and numerical models, still the paper
\cite{BSt} highlighted the desirability of an additional arbitrary
parameter (representing a rotational octupole moment) in the
analytic solutions that pretend to describe a generic NS. The
desired octupole parameter is naturally contained in the vacuum
sector of the MMR metric defined, within the framework of the
Ernst formalism \cite{Ern,Ern2} and Sibgatullin's method
\cite{Sib}, by the axis data of the form \cite{MMR}\footnote{Note
that, compared to the original paper \cite{MMR}, we have
introduced a formal sign change $k\to-k$ in the axis data
(\ref{AD}).}
\be \E(\rho=0,z)\equiv
e(z)=\frac{(z-m-ia)(z+ib)+k}{(z+m-ia)(z+ib)+k}, \label{AD} \ee
where the four arbitrary real parameters $m$, $a$, $k$ and $b$ are
associated, respectively, with the mass, angular momentum,
mass-quadrupole and angular-momentum-octupole moments of the
source. In recent years, to a large extent due to the efforts of
Pappas and Apostolatos \cite{PAp3,PAp4,PAp5}, the corresponding
vacuum metric was shown to be the best analytical approximation to
the exterior gravitational field of any kind of unmagnetized
rotating NSs, and lately such an assessment has been strongly
supported by the discovery of various universal properties and
relations for NSs \cite{YYu,MCF}, accompanied by the explanation
of the physical mechanisms that may lie behind that universal
behavior \cite{YKP}. Nonetheless, despite all its notable
properties, the vacuum MMR solution defined by (\ref{AD}) has not
yet found a widespread use among the researchers, and in our
opinion this might partly be attributed to a very complicated form
that was given to it in the paper \cite{PAp4} where Pappas and
Apostolatos apparently did not use the most rational way of the
calculation of the metric function $\omega$ allowed by
Sibgatullin's method. Therefore, one of the main objectives of the
present paper will be giving a concise form for the whole MMR
4-parameter vacuum solution, together with the remarkably simple
expressions of its metric functions in the equatorial plane, what
we believe must make this solution very suitable for direct use in
astrophysical applications even by non-experts in the solution
generating techniques. Moreover, since it was already formally
shown in the paper \cite{PAp4} how different branches of the
two-soliton solution are related to the numerical results of Berti
and Stergioulas \cite{BSt}, in the present paper we are going to
reconsider this question in more detail, demonstrating in
particular how the concrete type of the solution (subextreme,
hyperextreme or mixed) can be read off from the numerical data of
the paper \cite{BSt} directly at the level of the multipole
moments, and we will speculate about a possible impact our
analysis might have on raising the physical status of hyperextreme
spacetimes.

Our paper is organized as follows. A concise form of the MMR
vacuum solution will be given in the next section, together with
our clarifying remarks on finding the metric function $\omega$ in
the vacuum and electrovacuum cases by means of Sibgatullin's
method; here we also obtain a simple representation of the
solution in the equatorial plane. In Sec.~III we first consider a
reparametrization of the axis data (\ref{AD}) in terms of four
multipole moments and then show how this permits one to study the
involvement of the subextreme and hyperextreme constituents in our
two-soliton solution when the numerical results of Berti and
Stergioulas \cite{BSt} are being used as matching data.
Furthermore, to make the interior structure of the MMR vacuum
solution more comprehensible to the reader, we will illustrate it
by comparing the latter solution with the extended two-soliton
solution from our paper \cite{MRu}. Here we also consider an exact
analog of the Hartle-Thorne approximate metric. Sec.~IV is devoted
to the discussion of possible physical implications of the results
obtained and to concluding remarks.

\section{The MMR 4-parameter vacuum solution}

Unlike the paper \cite{PAp4}, where the expression (\ref{AD}) was
presented by Pappas and Apostolatos as just an {\it ansatz}, we
first of all would like to remark that Eq.~(\ref{AD}) is not an
intuitive result or some fortunate guessing, but in reality it is
a direct outcome of thorough analysis of the multipole structure
of the extended two-soliton electrovac solution performed in the
paper \cite{MMR2}. The four arbitrary parameters in (\ref{AD})
correspond to four arbitrary Geroch-Hansen relativistic multipole
moments \cite{Ger,Hans}, and the form of these multipoles in terms
of the parameters $m$, $a$, $b$ and $k$ can be easily found from
(\ref{AD}) with the aid of the Fodor-Hoenselaers-Perj\'es
procedure \cite{FHP}, yielding \cite{MMR}
\be M_0=m, \quad M_2=-m(a^2+k), \quad J_1=ma, \quad
J_3=-m[a^3+k(2a-b)]. \label{mm} \ee
Here $M_0$ stands for the total mass, $M_2$ is the mass-quadrupole
moment, whereas $J_1$ is the total angular momentum of the source
and $J_3$ its angular-momentum-octupole moment. It is then clear
that the parameters $m$ and $a$ have the same physical
interpretation as in the Kerr solution \cite{Ker}, so that the two
remaining parameters $b$ and $k$ actually describe the deviations
of the MMR vacuum solution from the Kerr spacetime.

Although the expression of the complex potential $\E(\rho,z)$
defined by (\ref{AD}) and the corresponding metric functions
$f(\rho,z)$, $\gamma(\rho,z)$ and $\omega(\rho,z)$ entering the
Weyl-Papapetrou stationary axisymmetric line element
\be d s^2=f^{-1}[e^{2\gamma}(d\rho^2+d z^2)+\rho^2 d\varphi^2]-f(d
t-\omega d\varphi)^2, \label{WP} \ee
are readily obtainable from the 6-parameter electrovac solution of
the paper \cite{MMR} by just setting in it to zero the charge
parameter $q$ and the magnetic dipole parameter $c$,\footnote{Note
that the expression of the function $A$ in \cite{MMR} contained
some misprints that were later rectified in \cite{EMR}.} Pappas
and Apostolatos still opted in \cite{PAp4} for their own
rederivation of the vacuum solitonic solution from the general
formulas of the paper \cite{RMM}. However, they were seemingly
unaware of the important fact that in the pure vacuum case, when
the electromagnetic field is absent, the knowledge of only one of
the potentials $G$ or $H$ is really needed for the construction of
the metric coefficient $\omega$, not both of them. Actually, for
their purpose Pappas and Apostolatos should have better used the
formulas (2.1) of the paper \cite{MRu} which already take into
account the peculiarities of the vacuum case. Having in mind the
idea of improving the presentation of the vacuum MMR metric,
recently we have carefully revised our earlier work on the
extended two-soliton solutions, exploring in particular various
ways of writing the metric function $\omega$ of which we have
finally chosen the one that looked to us more attractive than the
others. However, before the presentation of the metric functions
of the MMR solution, below we first write down the form of the
Ernst potential $\E$ of the latter solution \cite{MMR,EMR}:
\bea \E&=&(A-B)/(A+B), \nonumber\\
A&=&\k_+^2\{[m^2(d-ab+2b^2)-(a-b)^2(d-ab-k)](R_+r_-+R_-r_+) \nonumber\\
&& -i\k_-[(a-b)(d-ab-k)+m^2b](R_+r_--R_-r_+)\} \nonumber\\
&&+\k_-^2\{[m^2(d+ab-2b^2)-(a-b)^2(d+ab+k)](R_+r_++R_-r_-) \nonumber\\
&& -i\k_+[(a-b)(d+ab+k)-m^2b](R_+r_+-R_-r_-)\}-4m^2kd(R_+R_-+r_+r_-), \nonumber\\
B&=&m\k_+\k_-\{d[\k_+\k_-(R_++R_-+r_++r_-)-(m^2-a^2+b^2)(R_++R_--r_+-r_-)]
\nonumber\\ &&+ibd[(\k_++\k_-)(R_+-R_-)+(\k_+-\k_-)(r_--r_+)]
\nonumber\\
&&+i[b(m^2-a^2)-ak][(\k_++\k_-)(r_+-r_-)+(\k_+-\k_-)(R_--R_+)]\},
\label{EP} \eea
where
\bea
R_\pm&=&\sqrt{\rho^2+\left(z\pm\frac{1}{2}\left(\k_++\k_-\right)\right)^2},
\quad
r_\pm=\sqrt{\rho^2+\left(z\pm\frac{1}{2}\left(\k_+-\k_-\right)\right)^2},
\nonumber\\ \k_\pm&=&\sqrt{m^2-a^2-b^2-2k\pm 2d}, \quad
d=\sqrt{(ab+k)^2-m^2b^2}. \label{Rr} \eea

It is not difficult to verify that on the upper part of the
symmetry axis ($\rho=0$, $z>{\rm Re}[(\k_++\k_-)/2]$) the above
potential $\E$ takes the form (\ref{AD}). Of course, it can also
be readily checked with a computer analytical program that $\E$
defined by (\ref{EP}) and (\ref{Rr}) satisfies identically the
Ernst equation
\be {\rm Re}(\E)(\E_{,\rho,\rho}+\rho^{-1}\E_{,\rho}+\E_{,z,z})
=\E_{,\rho}^2+\E_{,z}^2 \label{EE} \ee
(the comma in subindices denotes partial differentiation).

While the corresponding metric functions $f$ and $\gamma$ of the
vacuum MMR solution can be written in terms of the above
potentials $A$ and $B$ only, the form of the remaining function
$\omega$ involves the additional potential $G$ which we calculated
with the aid of the formulas of our paper \cite{MRu} in a bit more
rational way than this was done in the paper \cite{MMR}; the final
expressions for $f$, $\gamma$ and $\omega$ are the following:
\bea f&=&\frac{A\bar A-B\bar B}{(A+B)(\bar A+\bar B)}, \quad
e^{2\gamma}=\frac{A\bar A-B\bar B}{16d^2\k_+^4\k_-^4R_+R_-r_+r_-},
\quad \omega=2(a-b)-\frac{2{\rm Im}[G(\bar A+\bar B)]}{A\bar
A-B\bar B}, \nonumber\\ G&=&-zB
+\k_+^2\k_-[m^2(d-ab+2b^2)-(a-b)^2(d-ab-k)](R_-r_+-R_+r_-)
\nonumber\\
&&+\k_+\k_-^2[m^2(d+ab-2b^2)-(a-b)^2(d+ab+k)](R_-r_--R_+r_+)
+i\k_+^2\k_-^2\nonumber\\
&&\times\{d(a-b)(R_++R_-)(r_++r_-)
+[(a-b)(ab+k)-m^2b](R_+-R_-)(r_+-r_-) \nonumber\\
&&+mbd(R_++R_-+r_-+r_+)\}
+md\k_+\k_-\{\k_+(d+b^2+k)(R_--R_+-r_-+r_+)
\nonumber\\
&&+k_-(d-b^2-k) (R_+-R_--r_-+r_+) +i[(a-b)(ab+b^2+2k)-m^2b]
\nonumber\\ &&\times(R_++R_--r_--r_+) \} \label{mf} \eea
(a bar over a symbol means complex conjugation), and apparently
our way of expressing the metric function $\omega$ is by far
simpler than the formulas (14) and (B14)-(B20) of \cite{PAp4}.

It is worth noting that the function $\omega$ of the MMR vacuum
solution that follows straightforwardly from the formulas (3.11)
of \cite{MMR} (by setting $q=c=0$ and changing $k$ to $-k$) also
has quite a reasonable form somehow overlooked by Pappas and
Apostolatos, namely,
\bea \omega&=&-\frac{2{\rm Im}[m\bar L(A+B)]}{A\bar A-B\bar B},
\nonumber\\
L&=&\k_+^2\k_-^2[d(z-ia)(R_++R_-+r_++r_-)+ibm(R_+-R_-)(r_+-r_-)]
\nonumber\\
&&-imk(a-b)[4d(R_+R_-+r_+r_-)-\k_+^2(R_+r_-+R_-r_+)
+\k_-^2(R_+r_++R_-r_-)] \nonumber\\
&&+\k_+\k_-\{m[\k_+(d+b^2)(R_+r_--R_-r_+) +\k_-(d-b^2)(R_+r_+-R_-r_-)]
\nonumber\\
&&+d[(z-ia)(a^2-b^2-m^2)-2ik(a-b)](R_++R_--r_+-r_-)
\nonumber\\
&&+d(ab+k+ibz)[(\k_++\k_-)(R_+-R_-)+ (\k_+-\k_-)(r_--r_+)] \nonumber\\
&&+[k^2+abk+(a+iz)(a^2b+ak-bm^2)] [(\k_++\k_-)(r_--r_+)
\nonumber\\ &&+(\k_+-\k_-)(R_+-R_-)] \}, \label{om} \eea
where the factor 2 in the above expression of $\omega$ reflects
the fact that the contributions of the terms involving the
potentials $E$ and $L$ in the formula (3.11) of \cite{MMR} for
$\omega$ are identical in the absence of electromagnetic field, so
that only one of these potentials is needed in such case for the
construction of the field $\omega$. Nonetheless, we ourselves
still incline to the form (\ref{mf}) for $\omega$ because, on the
one hand, it is slightly more concise than the expression
(\ref{om}) and, on the other hand, we used it for elaborating a
nice representation of the MMR vacuum metric in the equatorial
plane that will be considered below.

Since in many practical astrophysical applications of the NS
models (the geodesic motion of test particles, the existence of
innermost stable circular orbits, energy release on the surface of
a NS by accreting matter, etc.) the analysis is usually restricted
to the equatorial plane of the exterior field, it is desirable to
have compact analytical expressions of all metric coefficients for
this important special domain defined in cylindrical coordinates
as $z=0$, $\rho\ge0$. Let us note that a very concise
``equatorial'' form for the 3-parameter solution discussed in the
papers \cite{SSu} was found by Sibgatullin and Sunyaev (see
Eqs.~(26) in the first paper of Ref.~\cite{SSu}), and its
knowledge was sufficient for being able to study extensively
various physical properties of NSs. Though the additional
parameter $b$ in the vacuum MMR 4-parameter solution apparently
complicates the general form of the metrical fields compared to
its particular 3-parameter specialization considered by
Sibgatullin and Sunyaev, we notwithstanding have been able to
obtain an analogous very simple representation for our more
general solution in the equatorial plane, and this required, apart
from just setting $z=0$ in the formulas (\ref{EP}), (\ref{Rr}) and
(\ref{mf}), some additional algebraic manipulations that
eventually led us to the following elegant result:
\bea f&=&\frac{A-B}{A+B}, \quad e^{2\gamma}=\frac{A^2-B^2}{({\rm
r}_++{\rm r}_-)^4{\rm r}_+^2{\rm r}_-^2}, \quad
\omega=-\frac{2mW}{A-B}, \nonumber\\ A&=&({\rm r}_++{\rm
r}_-)^2{\rm r}_+{\rm r}_--m^2k, \nonumber\\ B&=&m({\rm r}_++{\rm
r}_-) ({\rm r}_+{\rm r}_-+\rho^2-b^2-k),
\nonumber\\
W&=&({\rm r}_++{\rm r}_-)[a({\rm r}_+{\rm r}_-+\rho^2-b^2)-bk]+mk(a-b), \nonumber\\
{\rm r}_\pm&=&\sqrt{\rho^2+\frac{1}{4}\left(\k_+\pm\k_-\right)^2}.
\label{mfz0} \eea

The above formulas can also be rewritten in dimensionless form by
introducing
\be j=\frac{a}{m}, \quad  \beta=\frac{b}{m}, \quad
\k=\frac{k}{m^2}, \quad r=\frac{\rho^2}{m^2}, \quad {\tt
r}_\pm=\frac{\rm r_\pm}{m}, \label{rep} \ee
and thus yielding
\bea f&=&\frac{{\mathcal A}-{\mathcal B}}{{\mathcal A}+{\mathcal
B}}, \quad e^{2\gamma}=\frac{{\mathcal A}^2-{\mathcal B}^2}{({\tt
r}_++{\tt r}_-)^4{\tt r}_+^2{\tt r}_-^2}, \quad \frac{\omega}{m}=
-\frac{2{\mathcal W}}{{\mathcal A}-{\mathcal B}}, \nonumber\\
{\mathcal A}&=&({\tt r}_++{\tt r}_-)^2
{\tt r}_+{\tt r}_--\k, \nonumber\\
{\mathcal B}&=&({\tt r}_++{\tt r}_-) ({\tt r}_+{\tt
r}_-+r-\beta^2-\k),
\nonumber\\
{\mathcal W}&=&({\tt r}_++{\tt r}_-)[j({\tt r}_+ {\tt
r}_-+r-\beta^2)-\beta\k]+\k(j-\beta), \label{mfz0r} \eea
with ${\tt r}_\pm$ having the explicit form
\bea {\tt
r}_\pm&=&\sqrt{r+\frac{1}{4}\left(\sqrt{1-j^2-\beta^2-2\k+2\delta}
\pm\sqrt{1-j^2-\beta^2-2\k-2\delta}\right)^2}, \nonumber\\
\delta&\equiv&\sqrt{(\k+j\beta)^2-\beta^2}. \label{rpm} \eea
The expressions obtained by Sibgatullin and Sunyaev for the
3-parameter quadrupole solution in the equatorial plane are
recovered from (\ref{mfz0r}) and (\ref{rpm}) by putting $\beta=0$,
$\delta=\k$ and remembering that our $\k$ is their $b$.

Therefore, we have obtained the desired representation of the
vacuum MMR solution and its form in the equatorial plane. We now
turn to the discussion of the multipole structure of this solution
in relation with the type of the objects that may formally be
identified as its sources.

\section{Relationships between the MMR solution, EDK solution and
numerical solutions of Berti and Stergioulas}

The main advantage of the extended multi-soliton solutions
\cite{RMM,MRu} that were constructed by means of Sibgatullin's
method, and to which belongs in particular the MMR solution, is
that the parameters they contain correspond to arbitrary multipole
moments, so that in the pure vacuum case which is of interest to
us in this paper they naturally describe in a unified manner
arbitrary combinations of the subextreme and hyperextreme Kerr-NUT
constituents \cite{DNe}. As it follows from (\ref{mm}), the four
parameters in the axis data (\ref{AD}) do correspond to four
multipole moments, which means that (\ref{AD}) is reparametrizable
in terms of the multipole moments alone. This can be done either
by inverting formulas (\ref{mm}), namely,
\be m=M_0, \quad a=\frac{J_1}{M_0}, \quad
k=-\frac{M_2}{M_0}-\frac{J_1^2}{M_0^2}, \quad
b=\frac{-M_0J_3+M_2J_1}{M_0M_2+J_1^2}+\frac{J_1}{M_0}, \label{inv}
\ee
with the subsequent substitution of (\ref{inv}) into (\ref{AD}),
or by using directly formula (2.8) of \cite{MMR} in which one has
to take into account that in the equatorially symmetric case the
complex quantities $m_n$, $n=\overline{1,4}$, are related to $M_n$
and $J_n$ by
\be m_0=M_0, \quad m_1=iJ_1, \quad m_2=M_2, \quad m_3=iJ_3.
\label{mMJ} \ee
However, since the problem of parametrizing the axis data of the
general multi-soliton vacuum solution in terms of the multipole
moments was solved in our paper \cite{MRu}, below we shall simply
use the determinantal expressions obtained there and apply them to
the $N=2$ case, with $m_n$ defined by (\ref{mMJ}). Then for $e(z)$
we get the following representation:
\be e(z)=\frac{e_-(z)}{e_+(z)}, \quad e_\pm(z)=(L_2)^{-1}\left|
\begin{array}{ccc}
z^2\pm M_0z\pm iJ_1 & M_2 & iJ_3 \\ z\pm M_0 & iJ_1 & M_2 \\
 1 & M_0 & iJ_1
\end{array}
\right|, \quad L_2=\left|
\begin{array}{cc}
iJ_1 & M_2 \\ M_0 & iJ_1
\end{array}
\right|, \label{ezMR} \ee
where $e_+(z)$ and $e_-(z)$ stand for  $R(z)$ and $P(z)$ of the
paper \cite{MRu}, respectively. Note that the ratio of two
quadratic polynomials in $z$ determining $e(z)$ in (\ref{ezMR})
degenerates to the ratio of two linear in $z$ polynomials when
$L_2\equiv -M_0M_2-J_1^2=0$, $M_0\ne0$, in which case the
resulting $e(z)$ just defines the Kerr solution.

We emphasize that the form (\ref{ezMR}) of $e(z)$ is fully
equivalent to the initial axis data (\ref{AD}) of the vacuum MMR
solution, and therefore (\ref{ezMR}) can be used to analyze the
structure of this solution on a par with (\ref{AD}). The main
benefit of having the representation (\ref{ezMR}) of $e(z)$ in
terms of the multipole moments consists in the possibility to use
it for establishing a straightforward correspondence between the
MMR solution and the numerical solutions of Berti and Stergioulas
\cite{BSt}, and in particular to answer an important question of
which types (subextreme or hyperextreme) of the Kerr-NUT sources
constituting the MMR solution correspond to each numerical
Berti-Stergioulas model defined by a concrete numerical set of
multipoles $\{M_0,J_1,M_2,J_3\}$. Recall that in Sibgatullin's
method, when the consideration is restricted to the pure vacuum
case, the roots $\a_n$ of the equation
\be e(z)+\bar e(z)=0 \label{ES} \ee
play a fundamental role as they enter in the ultimate expressions
of the metric functions, in this way introducing the parameters of
the initial axis data into the final formulas. For example, the
four roots of equation (\ref{ES}) with the MMR data (\ref{AD}) are
contained in the expressions (\ref{Rr}) for the functions $R_\pm$
and $r_\pm$ as the constant quantities
\be \a_1=-\a_4=\frac{1}{2}(\k_++\k_-), \quad
\a_2=-\a_3=\frac{1}{2}(\k_+-\k_-), \label{alf} \ee
and it follows from the mathematical structure of (\ref{ES}), in
the case of rational axis data leading to polynomial equations
with real coefficients, that the roots $\a_n$ can take either real
values or occur in complex conjugate pairs. Then a pair of real
$\a$'s in the MMR vacuum solution defines a subextreme Kerr-NUT
constituent, while a pair of complex conjugate $\a$'s determines a
hyperextreme Kerr-NUT constituent. In application to the
reparametrized axis data (\ref{ezMR}) this means that the
substitution of (\ref{ezMR}) into (\ref{ES}), yielding the
biquadratic equation
\bea z^4+\left(M_0^2+\frac{M_2^2+2J_1J_3}{M_0M_2+J_1^2}
+\frac{M_0(M_2^3+2M_2J_1J_3-M_0J_3^2}{(M_0M_2+J_1^2)^2}\right)z^2
\nonumber\\ +\left(\frac{M_2^2+J_1J_3}{M_0M_2+J_1^2}\right)^2
-\left(J_1+\frac{M_0(M_2J_1-M_0J_3)}{M_0M_2+J_1^2}\right)^2=0,
\label{eqS} \eea
permits us, by solving (\ref{eqS}) for any given set of multipoles
$\{M_0,J_1,M_2,J_3\}$ from \cite{BSt} determining a particular
Berti-Stergioulas numerical model of a NS, to obtain the
corresponding set of four $\a_n$ and hence answer at once the
question about the type of the two Kerr-NUT sources constituting
that concrete physically realistic numerical model. It is clear
that the roots of equation (\ref{eqS}) can give rise to the
following four generic types of the two-soliton configurations for
the NS models: type I configurations defined by four real-valued
$\a$'s that correspond to two subextreme Kerr-NUT sources; type II
configurations defined by two real $\a$'s and a pair of complex
conjugate $\a$'s describing one subextreme and one hyperextreme
Kerr-NUT constituents; type III configurations defined by two
pairs of complex conjugate $\a$'s determining a pair of identical
hyperextreme Kerr-NUT constituents located above and below the
equatorial plane; and lastly, type IV configurations defined by
four pure imaginary $\a$'s describing a pair of non-identical
overlapping Kerr-NUT constituents located in the equatorial plane
(see Fig.~1). In what follows it will be interesting to see
whether matching of the Berti-Stergioulas numerical solutions for
NSs to the MMR vacuum solution involves all four sectors of the
two-soliton solution or only part of them.

At this point, we find it instructive to rectify one imprecise
characteristic given by Pappas and Apostolatos in \cite{PAp4} to
the paper \cite{RMM}, according to which the latter paper only
presents an ``algorithm'' for the construction of exact solutions,
requiring (in the pure vacuum case) resolution of equation
(\ref{ES}) for some prescribed axis data. In reality, the
``algorithm'' mentioned by Pappas and Apostolatos is just
Sibgatullin's method itself, while the main output of the paper
\cite{RMM} is of course the $6N$-parameter electrovac metric in
which the quantities $\a_n$, $n=\overline{1,2N}$, are contained as
{\it arbitrary} parameters. In \cite{RMM} it was revealed for the
first time that the analytical resolution of the algebraic
equation for the axis data, which had been always considered a
necessary step in Sibgatullin's method but in practice had been
restricting the method's coverage mainly by the two-body systems
(since already the generic three-body configuration leads to the
algebraic equation of the sixth order), is not really needed, as
the roots of that algebraic equation themselves can be introduced
as arbitrary parameters into the multi-soliton solutions instead
of certain parameters of the axis data. As a matter of fact, it is
precisely the paper \cite{RMM} that had converted the method
initially considered as limited by various researchers (including
Sibgatullin himself) into a powerful tool for the analytical study
of the multi-black-hole configurations. In order to illustrate the
above said, let us see how the MMR vacuum solution is contained in
the extended double-Kerr (EDK) solution whose Ernst complex
potential $\E$ has the form \cite{MRu,MRS}
\bea \E&=&E_-/E_+, \nonumber\\ E_\mp&=&(\a_2-\a_3)(\a_1-\a_4)
[X_1r_1-X_2r_2\mp(\a_1-\a_2)][X_3r_3-X_4r_4\mp(\a_3-\a_4)]
\nonumber\\ &&-(\a_1-\a_2)(\a_3-\a_4)
[X_2r_2-X_3r_3\mp(\a_2-\a_3)][X_1r_1-X_4r_4\mp(\a_1-\a_4)],
\nonumber\\ r_n&=&\sqrt{\rho^2+(z-\a_n)^2}, \quad
X_n=\frac{(\a_n-\bar\beta_1)(\a_n-\bar\beta_2)}
{(\a_n-\beta_1)(\a_n-\beta_2)}, \label{EPMR} \eea
where $\beta_1$ and $\beta_2$ are arbitrary complex constants, and
$\a_n$, $n=\overline{1,4}$, can take arbitrary real values or form
complex conjugate pairs. Without lack of generality, the
parameters $\a_n$ can be assigned the order ${\rm Re}(\a_1)\ge{\rm
Re}(\a_2)\ge{\rm Re}(\a_3)\ge{\rm Re}(\a_4)$, and it can be easily
verified that on the upper part of the symmetry axis ($\rho=0$,
$z>{\rm Re}(\a_1)$) the equation (\ref{ES}) corresponding to the
EDK solution (\ref{EPMR}) rewrites as
\be \frac{2\prod_{n=1}^4(z-\a_n)}
{\prod_{l=1}^2(z-\beta_l)(z-\bar\beta_l)}=0 \label{ESEDK} \ee
for any combination of real and complex conjugate $\a$'s, whence
it follows immediately that $\a_n$ are indeed the roots of
equation (\ref{ES}) from Sibgatullin's method, although we did not
have to find them by solving the latter equation.

It was already pointed out in \cite{MRS} that the equatorially
symmetric subfamily of the EDK solution is defined by the
conditions
\bea \a_1+\a_4=\a_2+\a_3=0, \nonumber\\ X_1X_4=X_2X_3=-1,
\label{sym} \eea
so that formulas (\ref{EPMR}) and (\ref{sym}) combined together
are equivalent to the MMR vacuum solution and, importantly, they
permit us to give below a systematic description of all sectors of
this solution depicted in Fig.~1.

($a$) {\it Type I configurations in MMR solution}. When all $\a_n$
are real-valued, the corresponding constant quantities $X_n$, as
can be readily seen from their definition in (\ref{EPMR}), satisfy
the relations $X_n\bar X_n=1$, hence being the unit complex
constants. In this case, accounting for (\ref{sym}), the solution
can be parametrized by the following four quantities,
\be \{\a_1,\a_2,X_1,X_2\}, \label{par1} \ee
the remaining parameters $\a_3$, $\a_4$, $X_3$ and $X_4$ being
determined in terms of these as
\be \a_3=-\a_2, \quad \a_4=-\a_1, \quad X_3=-1/X_2, \quad
X_4=-1/X_1. \label{a34} \ee
Both Kerr-NUT constituents defined by this branch are subextreme.

($b$) {\it Type II configurations in MMR solution}. In these
configurations composed of one subextreme and one hyperextreme
Kerr-NUT constituents, the constants $\a_1$ and $\a_4$ are
real-valued, while $\a_2$ and $\a_3$ are complex conjugate pure
imaginary parameters $\a_2=\bar\a_3=ip$, where $p$ is an arbitrary
real constant. Then it follows from (\ref{EPMR}) that
\be X_1\bar X_1=X_4\bar X_4=1,  \quad X_2\bar X_3=1, \label{X2}
\ee
thus identifying $X_1$ and $X_4$ as unit complex constants, while
$X_2$ and $X_3$, with account of (\ref{sym}), are pure imaginary
quantities. This sector of the MMR solution then can be
parametrized by the quantities
\be \{\a_1,\a_2(=ip),X_1,X_2(=iq)\}, \label{par2} \ee
where $q$ is an arbitrary real constant, and the remaining
parameters $\a_3$, $\a_4$, $X_3$, $X_4$ are related to the above
$\a$'s and $X$'s via the formulas (\ref{a34}), as in the previous
case.

($c$) {\it Type III configurations in MMR solution}. This case is
characterized by two pairs of complex conjugate $\a$'s,
$\a_1=\bar\a_2$ and $\a_3=\bar\a_4$, describing two hyperextreme
Kerr-NUT constituents separated by the equatorial plane, the
corresponding $X_n$ satisfying the relations
\be X_1\bar X_2=X_3\bar X_4=1. \label{X3} \ee
The additional conditions (\ref{sym}) of equatorial symmetry then
imply that this sector of the MMR solution can be parametrized by
only two arbitrary complex quantities
\be \{\a_1,X_1\}, \label{par3} \ee
all the rest parameters being expressible in terms of $\a_1$ and
$X_1$ only:
\be \a_2=-\a_3=\bar\a_1, \quad \a_4=-\a_1, \quad X_2=1/\bar X_1,
\quad X_3=-\bar X_1, \quad X_4=-1/X_1. \label{a234} \ee

($d$) {\it Type IV configurations in MMR solution}. In this case,
the two hyperextreme Kerr-NUT sources defined by four pure
imaginary $\a$'s lie entirely in the equatorial plane, and the
constants $X_n$, as it follows from (\ref{EPMR}), are related by
the equations
\be X_1\bar X_4=X_2\bar X_3=1. \label{X4} \ee
Moreover, conditions (\ref{sym}) reveal that all $X_n$ are pure
imaginary quantities, so that the two-soliton solution in this
case can be parametrized by two $\a$'s and two $X$'s:
\be \{\a_1(=ip_1),\a_2(=ip_2),X_1(=iq_1),X_2(=iq_2)\},
\label{par4} \ee
where $p_i$ and $q_i$, $i=1,2$, are arbitrary real parameters, the
remaining constant quantities $\a_3$, $\a_4$, $X_3$ and $X_4$
having the form
\be \a_3=-ip_2, \quad \a_4=-ip_1, \quad X_3=i/q_2, \quad
X_4=i/q_1. \label{X34} \ee

Note that all types of the configurations ($a$)--($d$) involve
four arbitrary real parameters, which suggests that all sectors of
the MMR solution might also be equally important from the physical
point of view. The above analysis of the possible combinations of
subextreme and hyperextreme Kerr-NUT objects described by the MMR
solution complements and extends the analogous analysis performed
by Pappas and Apostolatos directly in terms of the proper
parameters of that solution, and we believe that our consideration
permits one to see more clearly the interrelations between
different sectors of the extended two-soliton solution. In
particular, it is apparent that sector ($a$) of the MMR solution
is equivalent to the equatorially symmetric (``parallel angular
momentum'') subfamily pointed out by Oohara and Sato \cite{OSa}
within the usual double-Kerr solution of Kramer and Neugebauer
\cite{KNe} describing exclusively the subextreme Kerr-NUT
constituents; therefore, since both Oohara and Sato, and earlier
Kramer and Neugebauer themselves had indicated how the Kerr
black-hole metric is obtainable from the subextreme double-Kerr
solution, the affirmation made by Pappas and Apostolatos about the
absence of the Kerr metric among the configurations of type~I is
erroneous. As a matter of fact, the Kerr black hole spacetime is
definitely contained in the $b=0$ subfamily of the type~I MMR
solution.

Turning now to the numerical models of NSs constructed and
discussed by Berti and Stergioulas in \cite{BSt}, it should be
observed that their matching to the MMR vacuum solution needs all
four sectors of the latter solution. This can be seen already by
analyzing the first two sequences of numerical solutions from
table~1 of \cite{BSt} constructed with EOS A. In our Tables~I and
II we give the values of the multipole moments defining those
numerical solutions and the corresponding values of the parameters
$\a_1$ and $\a_2$ obtained by solving equation (\ref{eqS}) that
determine a respective sector (type) of the MMR solution. It is
really surprising that only two numerical models can be matched to
the type~I MMR analytic solution involving two subextreme Kerr-NUT
constituents, whereas seven models from Table~I and all nine
models from Table~II require sectors with at least one
hyperextreme Kerr-NUT constituent for matching to the MMR
solution. In addition, we have checked that the entire
supramassive sequence of ten models from table~1 of \cite{BSt}
matches to the type~II MMR solution. The situation is quite
similar in the case of the EOS AU numerical models of \cite{BSt}
too, whose type was investigated by Pappas and Apostolatos in
\cite{PAp4} using the improved values of the multipoles $M_2$ and
$J_3$,\footnote{In Tables I and II we have used the original data
from the paper \cite{BSt}.} and for which only one of the total
thirty models matches to the type~I MMR solution. This clearly
shows, on the one hand, that the subextreme double-Kerr solution
is not appropriate for approximating a generic NS model and, on
the other hand, that the hyperextreme Kerr-NUT objects are
physically important and play a fundamental role in the
description of the exterior field of NSs.

Taking into account the historical importance of the Hartle-Thorne
approximate solution \cite{HTh} for astrophysics, it would
certainly be of interest to conclude this section by considering
an exact analog of this well-known 3-parameter spacetime.
Fortunately, such an exact analog can be easily identified thanks
to the important fact established by Pappas and Apostolatos in
\cite{PAp4} -- the octupole rotational moment of the Hartle-Thorne
solution is equal to zero. This means that the particular
3-parameter specialization of the MMR vacuum solution with $J_3=0$
represents the desired exact ``Hartle-Thorne'' spacetime. From
(\ref{mm}) it follows that $J_3$ vanishes when $b=a(a^2+2k)/k$, so
that the substitution of this value of $b$ into the equations
(\ref{EP}), (\ref{Rr}) and (\ref{mf}) formally solves the problem
of describing the desired exact solution in explicit form.
Nonetheless, we still find it advantageous to rewrite the MMR
solution in terms of the dimensionless multipole moments $j$, $q$,
$s$ related to $M_0$, $J_1$, $M_2$ and $J_3$ by the formulas
\be j=\frac{J_1}{M_0^2}, \quad q=\frac{M_2}{M_0^3}, \quad
s=\frac{J_3}{M_0^4}, \label{dmm} \ee
in order to make the passage to the ``Hartle-Thorne'' subcase
really trivial. The reparametrized MMR solution worked out with
the help of the results of the paper \cite{MRu} then can be
presented (after additionally setting $M_0=m$) in the following
final form:
\bea \E&=&\frac{A-B}{A+B}, \quad f=\frac{A\bar A-B\bar
B}{(A+B)(\bar A+\bar B)}, \quad e^{2\gamma}=\frac{(j^2+q)^2(A\bar
A-B\bar
B)}{16\delta^2\k_+^4\k_-^4R_+R_-r_+r_-}, \nonumber\\
\omega&=&\frac{2m(s-jq)}{j^2+q}-\frac{2{\rm Im}[G(\bar A+\bar
B)]}{A\bar A-B\bar B}, \nonumber\\ A&=&\k_-^2(R_+-r_-)(R_--r_+)
-\k_+^2(R_+-r_+)(R_--r_-), \nonumber\\
B&=&m\k_+\k_-[(\k_++\k_-)(r_+-r_-) +(\k_+-\k_-)(R_--R_+)],
\nonumber\\ G&=&-zB +m\k_+\k_-[\k_-(R_+r_+-R_-r_-)
+\k_+(R_-r_+-R_+r_-)\nonumber\\ &&-2m\delta(R_++R_--r_--r_+)],
\label{mfn} \eea
where
\bea R_\pm&=&\frac{1-Y_\pm}{1+Y_\pm}
\sqrt{\rho^2+\left(z\pm\frac{m}{2}\left(\k_++\k_-\right)\right)^2},
\quad r_\pm=\frac{1-y_\pm}{1+y_\pm}
\sqrt{\rho^2+\left(z\pm\frac{m}{2}\left(\k_+-\k_-\right)\right)^2},
\nonumber\\
\k_\pm&=&\sqrt{1+j^2+2q-\left(j+\frac{jq-s}{j^2+q}\right)^2\pm
2\delta}, \quad \delta=\sqrt{\left(\frac{q^2+js}{j^2+q}\right)^2
-\left(j+\frac{jq-s}{j^2+q}\right)^2}, \label{Rrn} \eea
and
\be Y_\pm=\frac{\dfrac{i(jq-s)}{j^2+q}[\pm(\k_++\k_-)-2]-2ij}
{\dfrac{(jq-s)^2}{(j^2+q)^2}\pm(\k_++\k_-)-\k_+\k_--1}, \quad
y_\pm=\frac{\dfrac{i(jq-s)}{j^2+q}[\pm(\k_+-\k_-)-2]-2ij}
{\dfrac{(jq-s)^2}{(j^2+q)^2}\pm(\k_+-\k_-)+\k_+\k_--1}.
\nonumber\\ \label{Yy} \ee
The ``Hartle-Thorne'' exact solution is then obtainable from the
above formulas (\ref{mfn})-(\ref{Yy}) by just putting $s=0$ in
them. Note that the functions $R_\pm$ and $r_\pm$ in (\ref{Rrn})
are defined in a slightly different way than in (\ref{Rr}).

\section{Discussion and conclusions}

It seems very likely that the recent research on the numerical and
analytical modeling of the geometry around NSs is able to produce
a real breakthrough in our understanding of the important role
that the hyperextreme sources play in general relativity. First of
all, it is now clear that the inequality $j<1$, defining the
black-hole sector of the Kerr metric, in application to NSs may
define a variety of configurations comprised of the subextreme and
hyperextreme Kerr-NUT constituents. Indeed, although for all the
numerical Berti-Stergioulas models of NSs, as for all
astrophysical NSs in general, the inequality $j<1$ between the
total angular momentum and total mass holds, this inequality can
be satisfied, within the framework of the two-soliton solution not
only by the subextreme constituents, but by the hyperextreme
constituents too. Let us illustrate this with the following simple
calculation. Suppose, we have a system of $N$ identical corotating
hyperextreme Kerr sources characterized by the individual
dimensionless angular momentum/mass ratio $j_0>1$ each. Then it is
trivial to see that the corresponding ratio $j=J/M^2$ involving
the total angular momentum $J$ and total mass $M$ of the system
will be $j=j_0/N$, and hence for all $1<j_0<N$ we will have $j<1$,
so that this system formed exclusively by hyperextreme objects
will be seen as a subextreme source if we formally apply to it the
same criterion of subextremality as in the case of a single Kerr
black hole. Moreover, it is obvious that in the type~IV
two-soliton configurations considered in the previous section, the
hyperextreme sources lying in the equatorial plane can be
counterrotating and therefore have in principle arbitrarily large
individual spin parameters and at the same time give rise to the
system's $j<1$. Consequently, it would be plausible to suppose
that the question of whether or not the sources of certain exact
solutions describing the exterior geometry of physically realistic
NS models are subextreme or hyperextreme has no much sense,
especially in the context of the global NS models involving both
the exterior and interior solutions where the regions with
singularities do not show up at all; the only thing that really
matters is the capacity of exact solutions to describe all types
of sources.

Quite interestingly, the issue of the hyperextreme sources in the
NS models turns out to be intimately related to the universal
properties of NSs, in particular to a very attractive idea
proposed by Pappas and Apostolatos \cite{PAp5} that the exterior
gravitational field of a generic NS is determined by only four
arbitrary multipole moments. The latter idea was corroborated by
the explanation of possible mechanisms lying behind the situation
when the internal degrees of freedom encoded in different EOS of
NSs all lead to the exterior geometry described by just four
multipoles, and it looks to us fairly feasible. It is quite logic
to think that if a Kerr black black hole \cite{Ker} is described
by an equatorially symmetric 2-parameter one-soliton solution,
then the next-to-a-black-hole most compact and densest stellar
object, a NS, must be described by the next equatorially symmetric
soliton metric, the two-soliton one, possessing four arbitrary
parameters. If this is the case, the entire multipole structure of
NS models must be determined by only four arbitrary multipoles,
thus giving rise to a sort of a ``no-hair'' theorem for NSs. In
this respect it is worth noting that such a theorem, if correct,
would cover both the subextreme and hyperextreme sources
constituting the NS models, whereas the analogous ``no-hair''
theorem for uncharged black holes usually involves only the
subextreme sector of the Kerr solution. However, if, in
application to a Kerr black hole, we interpret the ``no-hair''
theorem as discriminating a spacetime geometry defined by only two
parameters -- mass and angular momentum -- then it is not quite
clear why we can not say for instance that ``a Kerr naked
singularity has no hair'' too. Indeed, the multipole structure of
the Kerr solution is concisely described by the formula~\cite{Her}
\be M_n+iJ_n=m(ia)^n, \label{HK} \ee
$m$ being the source's mass and $a$ its angular momentum per unit
mass, and it is obvious that the above formula equally applies to
the black-hole case ($m^2\ge a^2$) characterized by the presence
of an event horizon, and to the hyperextreme case ($m^2<a^2$)
characterized by the presence of a naked singularity. Therefore,
Kerr's multipole structure is independent of whether the event
horizon or a naked singularity is going to be formed during the
gravitational collapse, so that in both cases the same mechanism
must lie behind the structure's formation, the quantitative aspect
of the mass--angular-momentum relation playing a secondary role in
this process. Apparently, a key requirement this mechanism should
meet is to be able to trigger off the {\it final} gravitational
collapse.

It would probably be worth noting that during the late 70s and
early 80s of the last century the solution generating techniques
themselves had contributed a lot into drawing an artificial
dividing line between subextreme and hyperextreme spacetimes, when
the former spacetimes were generated with the aid of one ansatz
and the latter with the aid of a different one. The appearance of
Sibgatullin's integral method based on the general symmetry
transformation for the Ernst equations has changed that situation
drastically, as the extended multi-soliton solutions constructed
with its help contain parameters corresponding to arbitrary
multipole moments and hence describe in a unified manner both the
subextreme and hyperextreme sources. For example, in the case of
the Kerr solution, a starting point in Sibgatullin's method is the
axis data
\be \E(0,z)=\frac{z-m-ia}{z+m-ia}, \label{adK} \ee
where the real parameters $m$ and $a$ define two arbitrary
multipoles $M_0=m$ and $J_1=ma$, so that at the output we will
have
\bea \E(\rho,z)=\frac{(\k-ia)r_++(\k+ia)r_--2m\k}
{(\k-ia)r_++(\k+ia)r_-+2m\k}, \nonumber\\
r_\pm=\sqrt{\rho^2+(z\pm\k)^2}, \quad \k=\sqrt{m^2-a^2},
\label{Kerr} \eea
which is the Ernst potential of the Kerr spacetime that
automatically describes a black hole or a hyperextreme object,
depending on whether $m^2$ is greater or less than $a^2$. For
years, we have been calling for equal treatment of all types of
solutions arising within the extended solitonic spacetimes, and we
are really glad that the research of Berti and Stergioulas, and
that of of Pappas and Apostolatos, has finally produced a
convincing evidence, confirmed by the present paper, that the
exterior gravitational field of a generic NS can be appropriately
described exclusively by the extended soliton solution due to a
natural combination of its sub- and hyperextreme sectors.

Let us emphasize that we would be the first ones to admit that the
vacuum MMR solution, well suited for analytical approximation of
the geometry around NSs, might not be quite adequate for modeling
the exterior fields of other, less compact, stellar objects for
which a larger number of parameters representing arbitrary
multipole moments could be needed in the corresponding exact
solutions. In this regard, one can think about potential
importance the equatorially symmetric configurations of the
extended triple-Kerr \cite{MRM,PRS} and quadruple-Kerr \cite{HMM}
solutions might have for astrophysical applications in the future.

\section*{Acknowledgements}

This work was partially supported by CONACYT, Mexico, and by
Ministerio de Ciencia y Tecnolog\'\i a, Spain, under the Project
FIS2012-30926.

\newpage

\begin{figure}[htb]
\centerline{\epsfysize=55mm\epsffile{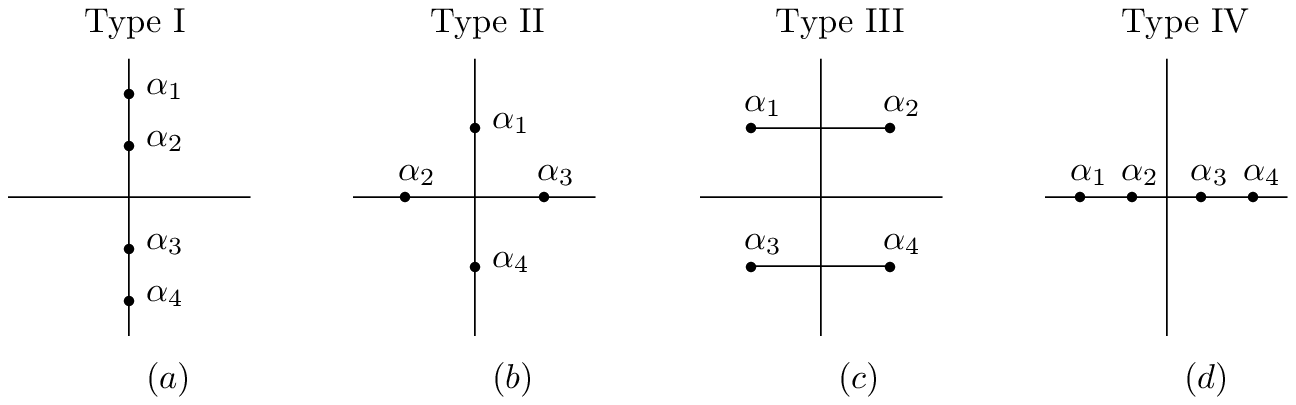}} \caption{Four
different sectors of the MMR solution determined by combinations
of the subextreme and hyperextreme Kerr-NUT constituents.}
\end{figure}

\newpage

\begin{table}[htb]
\caption{The multipole moments of the EOS A numerical models
(normal sequence) from table~1 of the paper \cite{BSt} and the
corresponding types of the MMR solution matching those
Berti-Stergioulas models.}
\begin{center}
\begin{tabular}{lllllll}
\hline \hline $M_0$ \hspace{1cm} & $J_1$ \hspace{1cm} & $M_2$
\hspace{1cm} & $J_3$
\hspace{1cm} & $\a_1$ \hspace{2cm} & $\a_2$ \hspace{1cm} & Type \\
\hline 2.074
& 0.8121  & -1.001 & -0.727 & 1.862 & 0.111 & I \\
2.081  & 1.327  & -2.656 & -3.158 & 1.386 & 0.484 & I \\
2.089  & 1.704  & -4.377 & -6.694 & 0.877-0.614i & $\bar\a_1$ & III \\
2.094 & 2.022 & -6.173 & -11.23 & 0.809-0.991i & $\bar\a_1$ & III \\
2.102  & 2.307 & -8.063 & -16.77 & 0.732-1.272i & $\bar\a_1$ & III \\
2.108 & 2.540 & -9.806 & -22.50 & 0.651-1.484i & $\bar\a_1$ & III \\
2.114 & 2.729 & -11.37 & -28.10 & 0.568-1.650i & $\bar\a_1$ & III \\
2.118 & 2.884 & -12.74 & -33.35 & 0.480-1.783i & $\bar\a_1$ & III \\
2.118 & 2.925 & -13.13 & -34.88 & 0.454-1.821i & $\bar\a_1$ & III \\
\hline \hline
\end{tabular}
\end{center}
\end{table}

\newpage

\begin{table}[htb]
\caption{The multipole moments of the EOS A numerical models
(second sequence) from table~1 of the paper \cite{BSt} and the
corresponding types of the MMR solution matching those
Berti-Stergioulas models.}
\begin{center}
\begin{tabular}{lllllll}
\hline \hline $M_0$ \hspace{1cm} & $J_1$ \hspace{1cm} & $M_2$
\hspace{1cm} & $J_3$
\hspace{1cm} & $\a_1$ \hspace{1cm} & $\a_2$ \hspace{1cm} & Type \\
\hline 2.453
& 0.9857  & -0.623 & -0.376 & 2.382 & -0.156i & II \\
2.462  & 1.573  & -1.657 & -1.637 & 2.265 & -0.265i & II \\
2.474  & 2.147  & -3.237 & -4.459 & 2.070 &  -0.377i & II \\
2.489 & 2.717 & -5.434 & -9.667 & 1.762 & -0.513i & II \\
2.50  & 3.143 & -7.529 & -15.72 & 1.412 & -0.652i & II \\
2.511 & 3.501 & -9.619 & -22.64 & 0.990 & -0.837i & II \\
2.521 & 3.746 & -11.31 & -28.76 & 0.542 & -1.028i & II \\
2.530 & 4.051 & -13.52 & -37.53 & -1.430i & -0.334i & IV \\
2.537 & 4.247 & -15.03 & -44.00 & -1.705i & -0.444i & IV \\
\hline \hline
\end{tabular}
\end{center}
\end{table}

\end{document}